\documentclass[%
preprint, 10pt,
 amsmath,amssymb,
 aps, prd
]{revtex4}
\usepackage{slashed}
\usepackage{mathrsfs}

\usepackage{graphicx}
\graphicspath{{pics/}}
\usepackage{dcolumn}
\usepackage{bm}
\usepackage{hyperref}

\usepackage[utf8]{inputenc}
\usepackage{color}
\allowdisplaybreaks[1]

\begin{document}
\title{Finite size effects in the free energy density for Abelian (anti-)self-dual gluon field in $SU(3)$ gluodynamics}

\author{Sergei N. Nedelko}
\email{nedelko@theor.jinr.ru}
\author{Vladimir E. Voronin}%
\email{voronin@theor.jinr.ru}
\affiliation{Bogoliubov Laboratory of Theoretical Physics, JINR, 141980 Dubna, Russia}%

\begin{abstract}
Finite-size effects in the free energy density for Abelian (anti-)self-dual gluon field are investigated within $SU(3)$ gluodynamics. 
In particular, the role of gluon quasi-zero modes is studied. The effective potential  is calculated within the framework of zeta function regularization for finite spherical four-dimensional region of radius $R$ in Euclidean space-time. In order to obtain the correct strong-field behavior of the effective potential which is  determined by the asymptotic freedom,  the quasi-zero gluon modes have to be treated beyond one-loop approximation in line with the argumentaion of  Leutwyler~\cite{Leutwyler:1980ma}. Conditions for appearance of the global minimum of the free energy density at finite nonzero values of both field strength and region size  are discussed.
\end{abstract}

\maketitle

\section{Introduction}
There exists certain evidence  that physical QCD vacuum can be rather efficiently  represented by the statistical ensemble of almost everywhere homogeneous Abelian (anti-)self-dual gluon field configurations. Indications come  from study of QCD effective action by various methods~\cite{Minkowski:1978fv,Leutwyler:1980ev,Leutwyler:1980ma,
Pagels:1978dd,Trottier:1992xp,Galilo:2010fn,Eichhorn:2010zc}
and application of the model of confinement, chiral symmetry breaking and hadronization in QCD, which is based on this representation of QCD vacuum, to calculation of hadron spectra, decay constants and formfactors~\cite{Efimov:1995uz,Kalloniatis:2001dw,
Kalloniatis:2003sa,Nedelko:2016gdk,Nedelko:2016vpj}.  In particular,  study of the effective action demonstrated that homogeneous Abelian (anti-)self-dual field appears to be a good candidate for a global minimum of the effective action~\cite{Leutwyler:1980ma,Eichhorn:2010zc}, and that there exist kink-like defects in the Abelian (anti-)self-dual homogeneous gluon background field~\cite{Galilo:2010fn} allowing one to explicitly represent general almost everywhere homogeneous  Abelian (anti-)self-dual field configurations in the form of
domain wall networks~\cite{Nedelko:2014sla}. Above-mentioned hadronization model implies not only the domain wall networks themselves, but also the existence of mean finite size of the regions with homogeneous field (domain bulk) in the typical domain wall network. This in turn requires certain mechanism that prevents domain size from infinite growth.  One potentially possible mechanism involves lower-dimentional topologically stable vortex, monopole and instanton-like configurations at the domain wall junctions which may stabilize the domain size. 
Another mechanism may relate to the  non-monotonic dependence of the domain energy on its size,   such that the energy minimum corresponds to certain finite size.  In fact, these scenarios may complement each other. 

In the present paper we investigate the second mechanism  for pure  gluodynamics, and observe that  gluon quasi-zero modes  can be crucially important for domain size stabilization. Namely, the free energy density (effective potential density) of the four-dimensional spherical domain with radius $R$ filled by the (anti-)self-dual Abelian field is calculated to the lowest non-vanishing order in gluon and ghost field fluctuations. The case of full QCD with quarks will be considered elsewhere. 

The free energy density  $F$ is defined by the Euclidean functional integral
\begin{equation}
\label{intF}
\exp\left(-V_R F(B,R)\right)=N\int_\mathscr{Q}\mathscr{D} Q\int_\mathscr{C} \mathscr{D} C \mathscr{D} C^\dagger\exp\left\{-\int_{V_R}d^4x\ \mathscr{L}(Q,C^\dagger,C,B)\right\}, 
\end{equation}
where $V_R$ is the volume of four-dimensional spherical region with radius $R$, and $\mathscr{L}(Q,C^\dagger,C,B)$ is the standard gauge-fixed Yang--Mills Lagrangian in the background gauge in the presence of the background gluon field 
\begin{gather*}
\breve{B}_\mu=\breve{n}B_\mu,\quad \breve{n}= T_3 \cos\xi+ T_8 \sin\xi,
\quad
B_\mu=-\frac{1}{2}B_{\mu\nu}x_\nu,
\\
B_{\mu\nu}=\pm\frac{1}{2}\varepsilon_{\mu\nu\alpha\beta}B_{\alpha\beta},\quad B_{\mu\alpha}B_{\nu\alpha}=B^2\delta_{\mu\nu},
\end{gather*}
where $\breve{B}_\mu$ stands for self-dual or anti-self-dual Abelian gluon field in adjoint  representation of $su(3)$, 
upper sign corresponds to self-dual field, and lower to anti-self-dual field. For finite Euclidean space-time region, the functional spaces $\mathscr{Q}$ and $\mathscr{C}$ contain 
gauge $Q$ and ghost $C$ fields subject to  Dirichlet boundary conditions 
\begin{equation}\label{dirichlet_bc}
\left. \breve  n Q_\mu(x)\right|_{x\in \partial V_R}=0, \quad \left. \breve n C(x)\right|_{x\in \partial V_R}=0, 
\end{equation}
 $V_R$ is the four-dimensional spherical region with boundary $\partial V_R$.
  Justification of this choice for the boundary conditions can be found in~\cite{Nedelko:2014sla,Kalloniatis:2001dw}. 
The normalization $N$ is chosen in such a way that the effective potential is equal
 to zero at vanishing background field strength.

The functional integral can be defined through decomposition of the gauge and ghost fields,
\begin{equation*}
Q(x)=\sum_{n}q_n Q^{(n)}(x),
\quad
C(x)=\sum_{n} c_n C_n(x),
\end{equation*}
over the basis in $\mathscr{Q}$ and $\mathscr{C}$  given by the  eigenfunctions of the corresponding differential operators,
\begin{align}
\left[-\breve D^2\delta_{\mu\nu}+2i\breve B_{\mu\nu}\right]Q^{(n)}_\nu=&\lambda_n Q^{(n)}_\mu
\nonumber\\
-\breve D^2 C_n = &\lambda_n C_n,
\label{eignvp}
\end{align} 
subject to boundary condition~\eqref{dirichlet_bc}.
Indices $n$ above are condensed ones: they include all relevant quantum numbers as described below.  It has to be stressed that the spectrum $(\lambda_n)$ is purely discrete for any  $R$ and any finite $B$. Integral~\eqref{intF}  takes the form
\begin{equation}
\label{intF1}
\exp\left(-V_R F(B,R)\right)=N\prod_{n,m,k}\int_{\mathscr{Q}} dq_n \int_{\mathscr{C}} d c_m d c_k^\dagger\exp\left\{-S(q,c,c^\dagger,B,R) \right\}.
\end{equation}
At finite $R$, all eigenvalues for gauge and ghost fields are positive, and the lowest (one-loop) order result for the free energy is 
obtained simply through determinants of the differential operators~\eqref{eignvp}. However, there will be neither regular strong-field nor infinite-volume limits for thus obtained free energy due to emergence of gluon zero modes (eigenmodes with zero eigenvalues) in these cases. 

Just ommiting the quasi-zero mode part of gauge field is not possible: if contribution of quasi-zero modes is neglected, the strong-field asymptotics of the free energy density does not comply  with asymptotic freedom~\cite{Leutwyler:1980ma}.

As it has been noticed by Leutwyler, contribution of the normal (nonzero) gluon and ghost modes  in the functional integral leads to a natural regularization of the zero modes. Already the lowest  correction due to normal modes  generate  a kind of ``effective mass'' term for zero modes and provides an appropriate Gaussian measure for integration over the  zero modes in the functional integral~\eqref{intF1}. 
Even though for finite-size region ($R<\infty$) zero modes turn into quasi-zero modes, the separate treatment of normal and quasi-zero modes is necessary. In the case of both strong-field and infinite-size limits, quasi-zero modes correctly tend to the corresponding zero modes. In order to provide the correct infinite-volume and strong-field limits of the effective potential,  one has to treat quasi-zero modes beyond one-loop approximation.

 In general, the effective ``mass'' for quasi-zero mode is expected to depend on  $R$. In this paper we analyze plausible forms of this behavior and its influence on the free energy density.
  Straightforward evaluation of this dependence will be given elsewhere. It is shown that incorporation of constant effective mass derived by Leutwyler for infinite volume leads to global minimum of the free energy density at nonzero finite $B$ and $R\to\infty$. 
This means that domain size would grow infinitely in pure gluodynamics. For effective ``mass''  vanishing at $R\to 0$, which is expected due to the rapid growth of all eigenvalues at small $R$, a global minimum at finite $B$ and $R$ arises.
Thus our main observation is that gluon quasi-zero modes  may be crucially important for domain size stabilization. 

The paper is organized as follows. 
Calculation of the free energy density is given in Section~\ref{section_ghosts_and_gluons}. Various scenarios for dependence of zero mode ``mass'' on domain size are analyzed in Section~\ref{effective mass}. The procedure of calculation of zeta-regularized determinants for ghost and gauge fields is described in detail in Appendix~\ref{ghost_gluon_detailed}. Auxiliary formulas are summarized in Appendices~\ref{Kummer_functions} and~\ref{asymptotic_expansion_appendix}. It should be noted that the calculational technique is one of the results by itself and is inseparable from the general physical content of the paper.

\section{Free energy density\label{section_ghosts_and_gluons}}

Classical part of free energy density is given by
\begin{equation*}
F^\textrm{cl}=B^2/g^2.
\end{equation*}
According to \eqref{intF1},
one-loop contribution to the free energy  is given by
\begin{equation}
\label{one-loop_via_det}
\delta U=V_R\delta F(B,R)=\frac{1}{2}\mathrm{Tr} \ln\Delta^\mathrm{gl}- 
\mathrm{Tr} \ln\Delta^\mathrm{gh}.
\end{equation}
Here $\Delta^\mathrm{gl}$ and  $\Delta^\mathrm{gh}$ stand for corresponding operators in \eqref{eignvp}.
We use analytical regularization
\begin{equation*}
\mathrm{Tr}\log\Delta=-\frac{d}{ds}\sum_j \left. \lambda_j^{-s}\vphantom{\frac11}\right|_{s=0}=-\left.\frac{d}{ds}\zeta(s)\right|_{s=0},
\end{equation*}
where $\lambda$ are eigenvalues of operator $\Delta$.
The method of computation of $\zeta(s)$ employed in the present study is summarized in Refs.~\cite{Bordag:1995gm,Kirsten:2001}.


The most straightforward calculation relates to the ghost fields. Operator $\Delta$ in this case is simply
\begin{equation*}
\Delta^\mathrm{gh}=-\breve{D}^2,\quad \breve{D}_\mu=\partial_\mu-i\breve{B}_\mu,
\end{equation*}
where $\breve{B}_\mu$ stands for self-dual or anti-self-dual Abelian gluon field in adjoint  representation of $su(3)$. The eigenvalues $\lambda$ of operator $\breve{D}^2$
are defined by equation~\cite{Kalloniatis:2001dw}
\begin{gather*}
M\left(\frac{k}{2}+1-m-\frac{\lambda^2}{2n_a B},k+2,\frac{n_a B R^2}{2}\right)=0,\ k=0,1,2,\dots,\ m=-\frac{k}{2},-\frac{k}{2}+1,\dots,\frac{k}{2}.
\end{gather*}
Here $n_a$ is the $a$-th nonzero eigenvalue of $\breve{n}$. The eigenmodes corresponding to zero eigenvalues of $\breve{n}$ do not depend on $B$ and hence do not contribute to the effective potential. Every solution of this equation with given $k,m$, radial number $r$ and $a$ is $k+1$-degenerate.
The spectrum is invariant with respect to $B\to -B$ which is evident if one performs Kummer transformation. In the limit $B\to 0$ equation for eigenvalues transforms to (see Appendix~\ref{Kummer_functions})
\begin{equation*}
(k+1)!\left(\frac{\lambda R}{2}\right)^{-k-1}J_{k+1}(\lambda R)=0.
\end{equation*}
The normalized one-loop contribution to effective potential is defined as
\begin{equation*}
\delta U^\mathrm{gh}=-\mathrm{Tr}\ln \frac{-\breve{D}^2}{ -\breve{\partial}^2}=-\sum_{kmr}\mathrm{Tr}_\mathrm{c}\ln\frac{\lambda_{kmr}^2(\breve{B},R)}{\lambda_{kmr}^2(0,R)}=\left.\frac{d}{ds}\zeta^\mathrm{gh}(s)\right|_{s=0},
\end{equation*}
where $\mathrm{Tr}_\mathrm{c}$ is trace with respect to color indices, and dimensionless quantities
\begin{equation*}
\lambda\to \lambda/\mu,\quad B\to B/\mu^2,\quad R\to R\mu
\end{equation*}
are introduced, $\mu$ is an arbitrary scale. 

\begin{figure}
\includegraphics[width=0.49\textwidth]{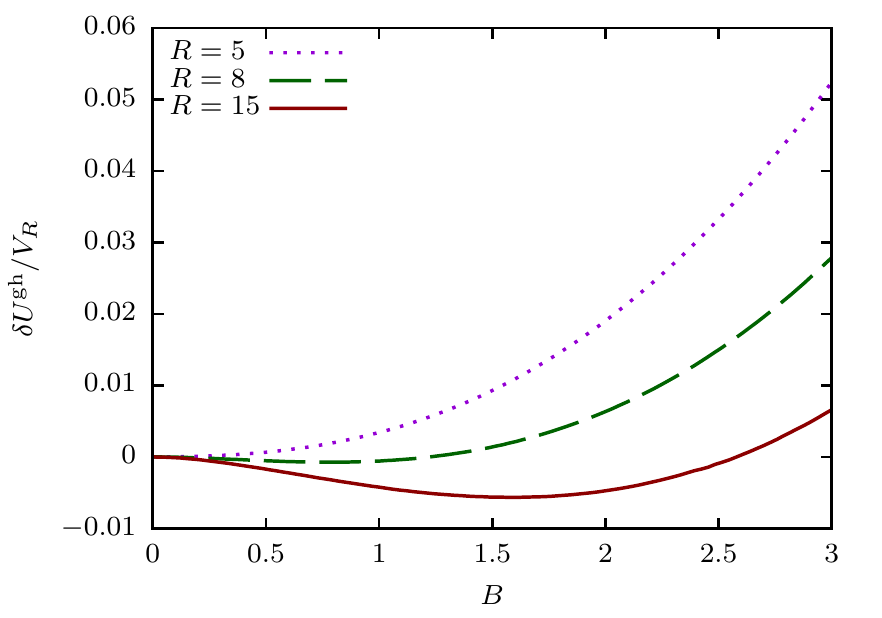}\hfill
\includegraphics[width=0.49\textwidth]{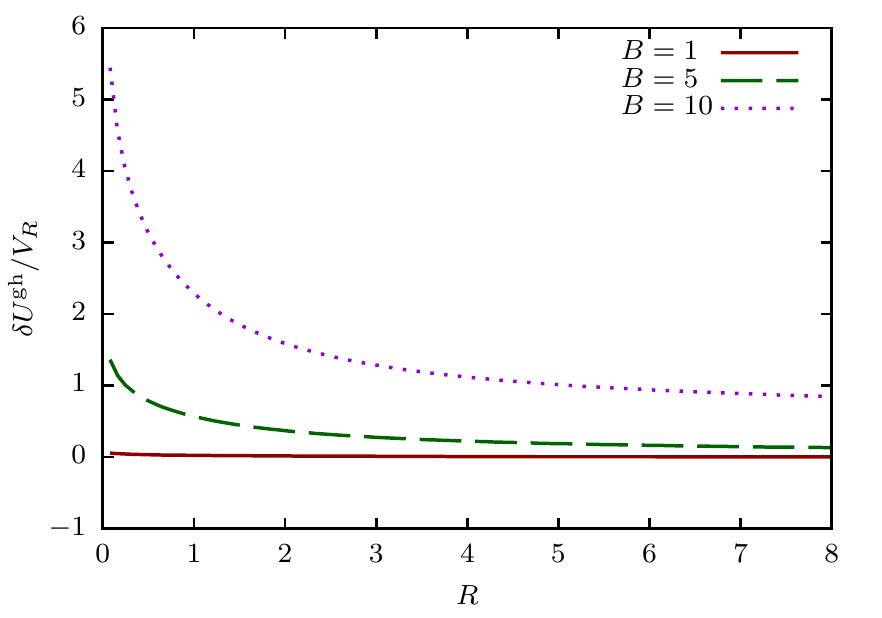}
\caption{Figures of free energy density of ghost field $\delta F^\textrm{gh}(B,R)=\delta U^\text{gh}/V_R$ at different values of $R$ and $B$.
There is a minimum versus $B$ if $R$ is sufficiently large. At large $R$, $\delta F^\textrm{gh}(B,R)$ tends to a constant.
\label{ghosts_figure}}
\end{figure}

Following Refs.~\cite{Bordag:1995gm,Kirsten:2001}, we can define zeta function as
\begin{gather*}
\zeta^\text{gh}(s)=\mathrm{Tr}_\mathrm{c}\frac{\sin \pi s}{\pi}\sum_{k=0}^\infty \sum_{m=-\frac{k}{2}}^\frac{k}{2}(k+1) \int_0^\infty \frac{dt}{t^{2s}}\frac{d}{dt}\Psi^\mathrm{gh}(k+1,m,t,\breve{B},R),\\
\Psi^\mathrm{gh}(k,m,t,\breve{B},R)=\log\frac{\exp\left(-\frac{\breve{B}R^2}{4}\right)M\left(\frac{k+1}{2}-m+\frac{t^2}{2\breve{B}},k+1,\frac{\breve{B}R^2}{2}\right)}{k!\left(\frac{t R}{2}\right)^{-k}I_{k}(t R)}.
\end{gather*}
This expression is still formal and divergent because regions of convergence of the integral and sums do not overlap. To make it finite at $s\to 0$, several terms of asymptotic decompositon of integrand  in $k\gg 1$  are added and subtracted
\begin{multline}
\label{ghosts_zeta_start}
\zeta^\text{gh}(s)=
\mathrm{Tr}_\mathrm{c}\left\{\frac{\sin\pi s}{\pi}\sum_{k=1}^\infty k^{1-2s} \int_0^\infty \frac{dt}{t^{2s}}\frac{d}{dt}\left[\sum_{m=-\frac{k-1}{2}}^\frac{k-1}{2}\Psi^\mathrm{gh}(k,m,kt ,\breve{B},R)-\sum_{i=0}^2  \frac{u_i^{\mathrm{gh}}(t,\breve{B},R)}{k^i}\right]\right.\\
\left.
+\frac{\sin\pi s}{\pi}\sum_{k=1}^\infty k^{1-2s} \int_0^\infty \frac{dt}{t^{2s}}\frac{d}{dt}\sum_{i=0}^2 \frac{u_i^{\mathrm{gh}}(t,\breve{B},R)}{k^i}
\right\}
\end{multline}
The first term in the curly brackets is analytical at $s\to 0$. The sums and integrals in the second term are expressed via analytical functions in the regions of $s$ where they converge, and analytically continued to $s\to 0$ in the complex plane (see Appendix~\ref{ghost_gluon_detailed} for details).
The final expression for $\delta U^\mathrm{gh}$ is
\begin{multline}
\label{ghosts_effpot}
\delta U^\mathrm{gh}(B,R)=-4\sum_{k=1}^\infty k \left[\sum_{m=-\frac{k-1}{2}}^\frac{k-1}{2}\Psi^\text{gh}(k,m,0,\frac{\sqrt{3}B}{2},R)-\frac{3}{4}B^2R^4\frac{1}{48}\left(1-\frac{1}{k}+\frac{1}{k^2}\right)\right]\\
+\frac{B^2R^4}{48}(2-3\gamma+3\log 2-3\log R)-\frac{3B^4R^8}{10240}.
\end{multline}
The first term is convergent sum that is computed numerically.
The contribution of ghosts to one-loop free energy density given by Eq.~\eqref{ghosts_effpot} is shown in Fig.~\ref{ghosts_figure}.


Completely analogous considerations for gluon field in Feynman gauge with
\begin{equation*}
\Delta_{\mu\nu}^\mathrm{gl}=-\breve{D}^2\delta_{\mu\nu}+2iB_{\mu\nu}
\end{equation*}
lead to equation
\begin{multline}
\label{gluons_effpot}
\delta U^\mathrm{gl}(B,R)=4\sum_{k=1}^\infty k \left[\sum_{m=-\frac{k-1}{2}}^\frac{k-1}{2}\Psi^\text{gl}(k,m,0,\frac{\sqrt{3}B}{2},R)-\frac{3}{4}B^2R^4\frac{1}{24}\left(1-\frac{1}{k}-\frac{5}{k^2}\right)\right]\\
-\frac{B^2R^4}{48}(31+30\gamma-30\log 2+30\log R)+\frac{3B^4R^8}{5120}.
\end{multline}
Contribution of $SU(3)$ gluodynamics to the effective potential of Abelian (anti-)self-dual gluon fields in finite volume is given by sum of Eqs.~\eqref{ghosts_effpot} and~\eqref{gluons_effpot}.

\begin{figure}
\includegraphics[width=0.49\textwidth]{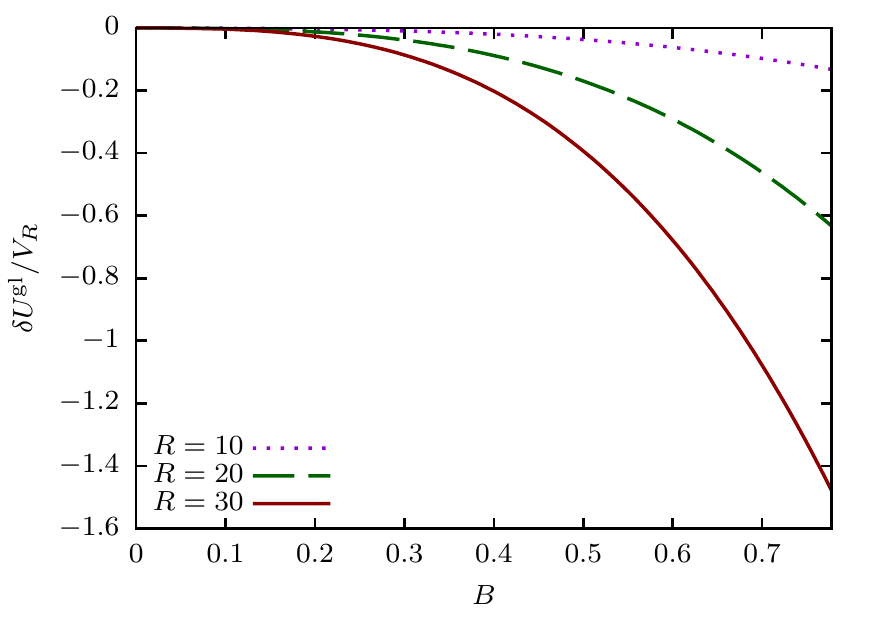}\hfill
\includegraphics[width=0.49\textwidth]{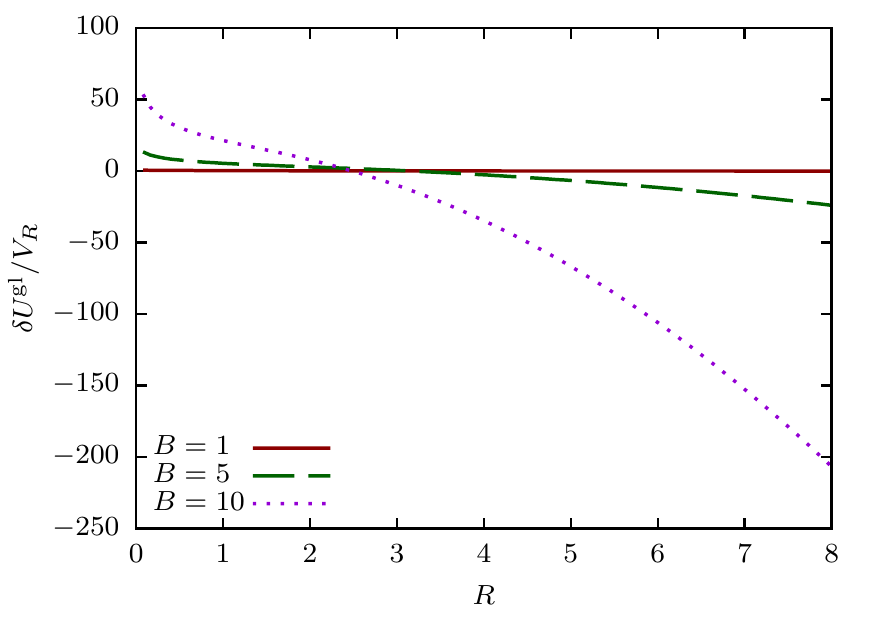}
\caption{Figures of free energy density of gluon field $\delta F^\textrm{gl}(B,R)=\delta U^\text{gl}/V_R$ (only one-loop contribution) at different values of $R$ and $B$. The strong-field asymptotics is incorrect, and there is no minimum with respect to $B$.  $\delta F^\textrm{gl}(B,R)$ does not have large-size limit.
\label{gluon_figure}}
\end{figure}
Free energy density of gluons $\delta F^\textrm{gl}(B,R)$ is shown in Fig.~\ref{gluon_figure}. It is easily seen that behavior at large $B$ and $R$ does not comply with predictions of renormalization group~\cite{Savvidy:1977as,Matinyan:1976mp} and results of calculations in infinite volume~\cite{Leutwyler:1980ma}. Free energy density  decreases at large $B$ and fixed $R$, and $\delta F^\textrm{gl}(B,R)$ does not approach a constant value at large $R$ and fixed $B$. This problem is a manifestation of quasi-zero gauge field eigenvalues that tend to zero as $BR^2\to\infty$.

\begin{figure}
\includegraphics[width=0.49\textwidth]{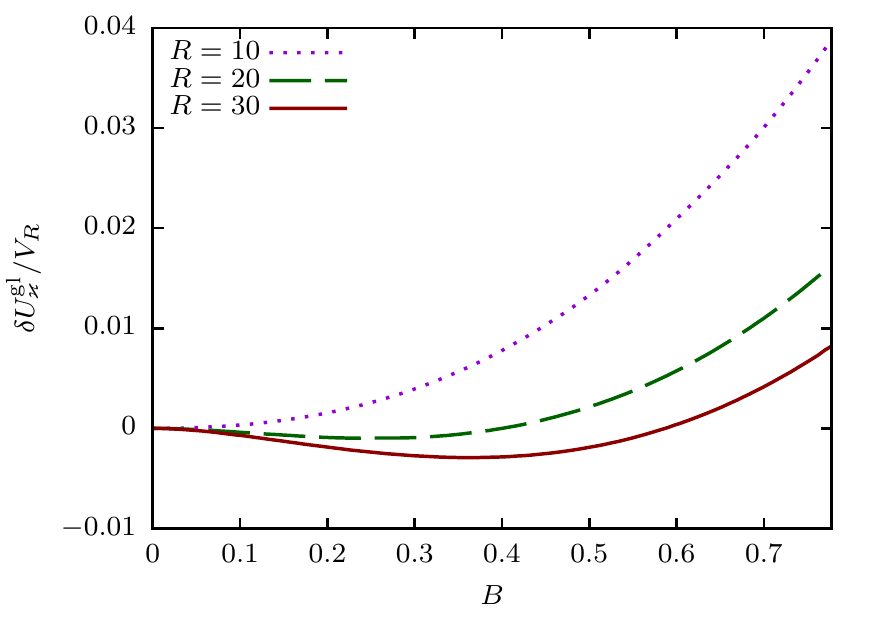}\hfill
\includegraphics[width=0.49\textwidth]{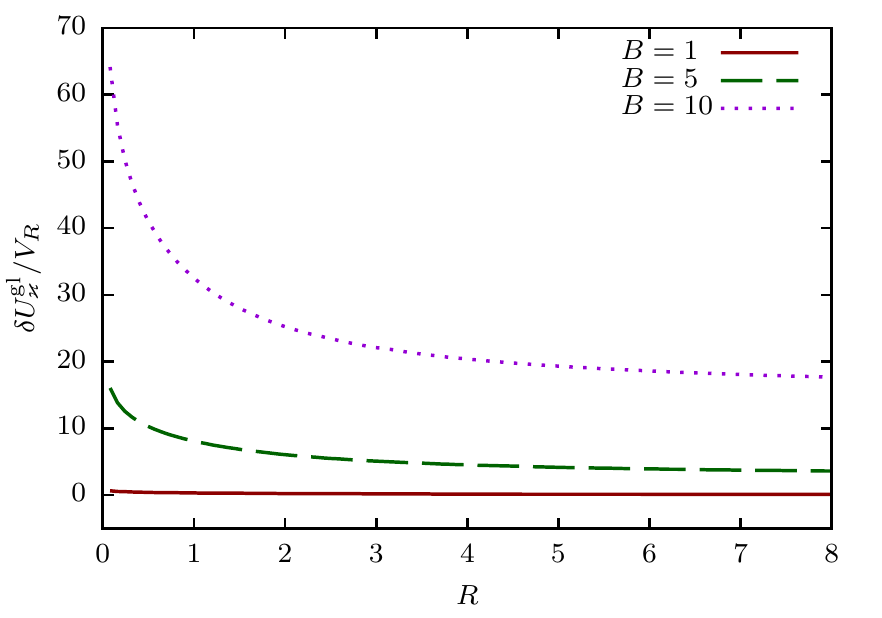}
\caption{Figures of free energy density $\delta F_\varkappa^\textrm{gh}(B,R)$ given by~\eqref{gluon_effpot_mass} at different values of $R$ and $B$ with $\varkappa=1$ incorporated. The correct strong-field asymptotics as well as large-size limit are restored, there is a minimum versus $B$ if $R$ is sufficiently large.
\label{gluon_effmass_figure}}
\end{figure}

If all eigenvalues are sufficiently large to provide Gaussian damping in the functional integral, then perturbation theory is applicable, and formula~\eqref{one-loop_via_det} is justified. The smaller the eigenvalues, the worse the one-loop approximation for contribution of quasi-zero modes. In the limit $BR^2\to\infty$ the lowest-order nonvanishing contribution due to quasi-zero modes comes not from the one-loop but from higher orders~\cite{Leutwyler:1980ma}. Thus, we have to take into account higher order  corrections in order to make one-loop effective potential reasonable at large $BR^2$. Calculations in the infinite volume is a guiding example. It was shown~\cite{Leutwyler:1980ma} that nonzero modes generate an effective ``mass term'' $\varkappa$ for zero modes. In terms of the present calculation, this mechanism leads to a shift in the quasi-zero eigenvalues $\lambda^2$ by a ``mass term'' 
\begin{equation*}
\lambda_\mathrm{eff}^2=\lambda^2+\varkappa B,
\end{equation*}
where $\varkappa$ is $B$-independent constant in the case of infinite volume, but in finite region it  depends on dimensionless variable $BR^2$. Calculation of $\varkappa$ in finite volume is rather complicated, but it is quite easy to estimate the potential effect which the quasi-zero mode ``mass'' may produce. If for present study we use constant $\varkappa$ that emerges in the case of infinite volume calculation, then incorporation of $\varkappa$ generates a term that restores behavior at large $BR^2$ (see Appendix~\ref{ghost_gluon_detailed} for details):
\begin{multline}
\label{gluon_effpot_mass}
\delta U_\varkappa^\mathrm{gl}(B,R)=
4\sum_{k=1}^\infty k \left[\sum_{m=-\frac{k-1}{2}}^\frac{k-1}{2}\Psi^\text{gl}(k,m,0,\frac{\sqrt{3}B}{2},R)-\frac{3}{4}B^2R^4\frac{1}{24}\left(1-\frac{1}{k}-\frac{5}{k^2}\right)\right]
\\
-\frac{B^2R^4}{48}(31+30\gamma-30\log 2+30\log R)+\frac{3B^4R^8}{5120}\\
+4\sum_{k=0}^\infty \left[(k+1)\log\frac{\lambda_{k\frac{k}{2}0}^2(B,R)+\varkappa B}{\lambda_{k\frac{k}{2}0}^2(B,R)}+(k+1)\log\frac{\lambda_{k\frac{-k}{2}0}^2(B,R)-\varkappa B}{\lambda_{k\frac{-k}{2}0}^2(B,R)}\right].
\end{multline}
Corresponding free energy density $\delta F_\varkappa^\textrm{gl}(B,R)=\delta U_\varkappa^\text{gh}/V_R$ is shown in Fig.~\ref{gluon_effmass_figure}.

\begin{figure}
\includegraphics[width=0.49\textwidth]{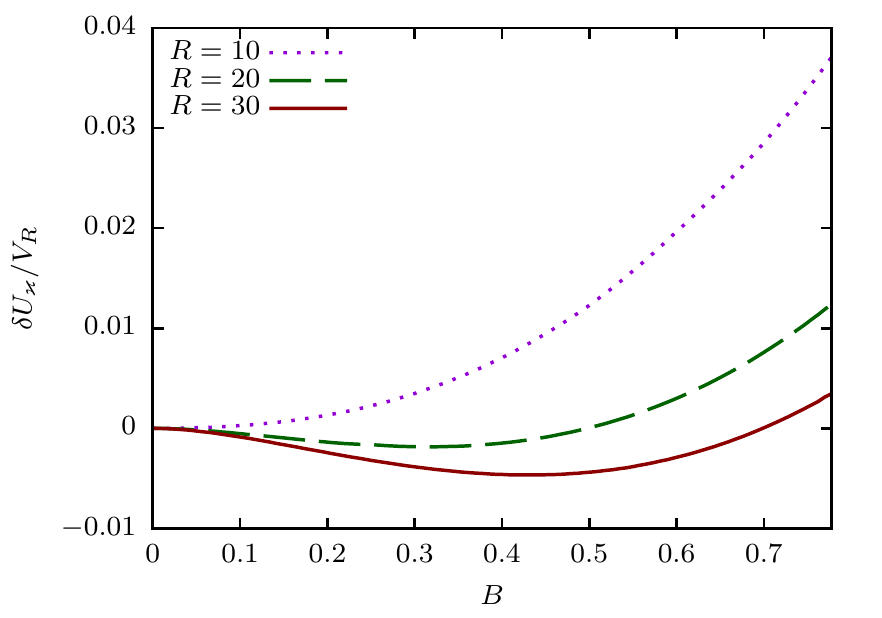}\hfill
\includegraphics[width=0.49\textwidth]{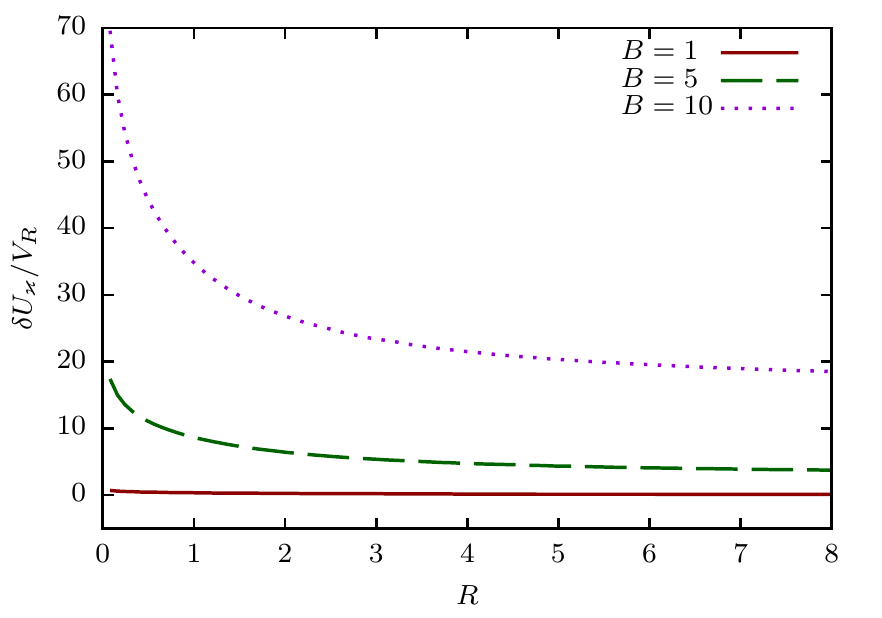}
\caption{Free energy density $\delta F_\varkappa(B,R)$ given by Eq.~\eqref{effpot_total_kappa} with $\varkappa=1$.\label{effpot_total_figure}}
\end{figure}

Combining one-loop contributions of gluons and ghosts together, one finds result of straightforward calculation (just calculation of the determinants including quasi-zero modes)
\begin{multline*}
\delta U(B,R)=\delta U^\mathrm{gl}(B,R)+\delta U^\mathrm{gh}(B,R)
=4\sum_{k=1}^\infty k \left[\sum_{m=-\frac{k-1}{2}}^\frac{k-1}{2}\Psi^\text{gl}(k,m,0,\frac{\sqrt{3}B}{2},R)-\frac{3}{4}B^2R^4\frac{1}{24}\left(1-\frac{1}{k}-\frac{5}{k^2}\right)\right]\\
-4\sum_{k=1}^\infty k \left[\sum_{m=-\frac{k-1}{2}}^\frac{k-1}{2}\Psi^\text{gh}(k,m,0,\frac{\sqrt{3}B}{2},R)-\frac{3}{4}B^2R^4\frac{1}{48}\left(1-\frac{1}{k}+\frac{1}{k^2}\right)\right]\\
-\frac{B^2R^4}{48}(29+33\gamma-33\log 2+33\log R)+\frac{3B^4R^8}{10120}.
\end{multline*}
Taking into account the higher-order  ``mass term'' $\varkappa$ for quasi-zero modes leads to
\begin{multline}
\label{effpot_total_kappa}
\delta U_\varkappa(B,R)=\delta U_\varkappa^\mathrm{gl}(B,R)+\delta U^\mathrm{gh}(B,R)
=4\sum_{k=1}^\infty k \left[\sum_{m=-\frac{k-1}{2}}^\frac{k-1}{2}\Psi^\text{gl}(k,m,0,\frac{\sqrt{3}B}{2},R)-\frac{3}{4}B^2R^4\frac{1}{24}\left(1-\frac{1}{k}-\frac{5}{k^2}\right)\right]\\
+4\sum_{k=0}^\infty \left[(k+1)\log\frac{\lambda_{\mathrm{gl},k\frac{k}{2}0}^2(B,R)+\varkappa B}{\lambda_{\mathrm{gl},k\frac{k}{2}0}^2(B,R)}+(k+1)\log\frac{\lambda_{\mathrm{gl},k\frac{-k}{2}0}^2(B,R)-\varkappa B}{\lambda_{\mathrm{gl},k\frac{-k}{2}0}^2(B,R)}\right]\\
-4\sum_{k=1}^\infty k \left[\sum_{m=-\frac{k-1}{2}}^\frac{k-1}{2}\Psi^\text{gh}(k,m,0,\frac{\sqrt{3}B}{2},R)-\frac{3}{4}B^2R^4\frac{1}{48}\left(1-\frac{1}{k}+\frac{1}{k^2}\right)\right]\\
-\frac{B^2R^4}{48}(29+33\gamma-33\log 2+33\log R)+\frac{3B^4R^8}{10120}.
\end{multline}
Total free energy density $F=\delta U_\varkappa(B,R)/V_R$ is shown in Fig.~\ref{effpot_total_figure}. The right-hand side plot indicates that infinite volume limit is well-defined as the free energy approaches constant value for any value of the field strength $B$. The left-hand side plot shows that there exists a minimum in $B$, and the field strength at the minimum decreases with decreasing $R$. In the limit $R\to\infty$, free energy density at strong field $B$  correctly approaches the known one-loop strong-field limit.  

The total free energy is independent of scale $\mu$~\cite{Wiesendanger:1993mw,Cognola:1993xy}:
\begin{equation*}
\mu \frac{d}{d\mu}U=0.
\end{equation*}
Combining $\delta U$ (only terms containing $\log R$ contribute to the above equation) with classical term, one obtains
\begin{gather*}
\mu\frac{d}{d\mu}\left[\frac{\pi^2B^2R^4}{2g^2}-\frac{11}{16}B^2R^4\log R\right]=0,\\
\mu\frac{d}{d\mu}g=-\frac{11g^3}{16\pi^2}.
\end{gather*}
That is, one-loop $\beta$-function of pure $SU(3)$ gluodynamics is recovered with the choice of boundary conditions adopted in the present study.

\section{Zero mode ``mass'' and domain size \label{effective mass}}

\begin{figure}
\includegraphics[width=0.49\textwidth]{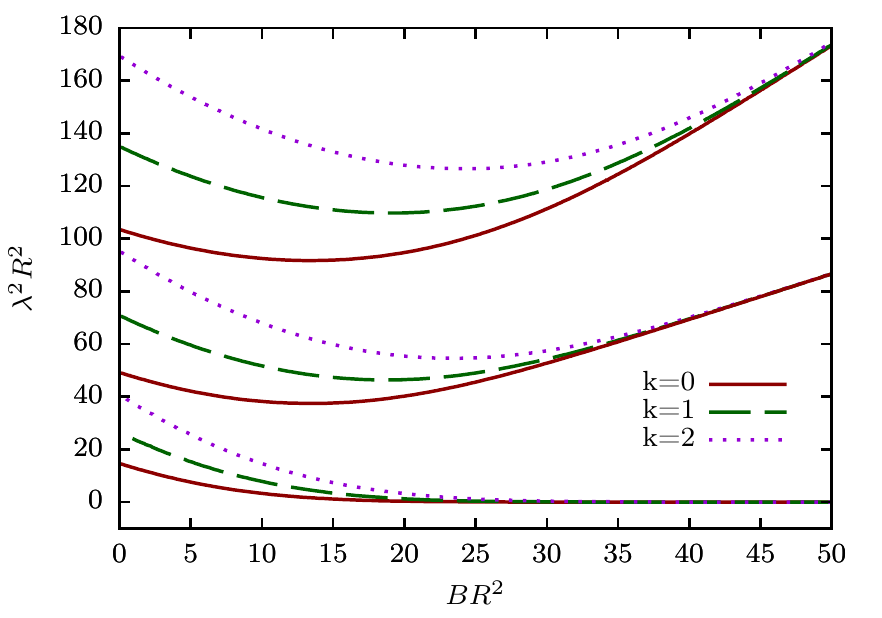}\hfill
\includegraphics[width=0.49\textwidth]{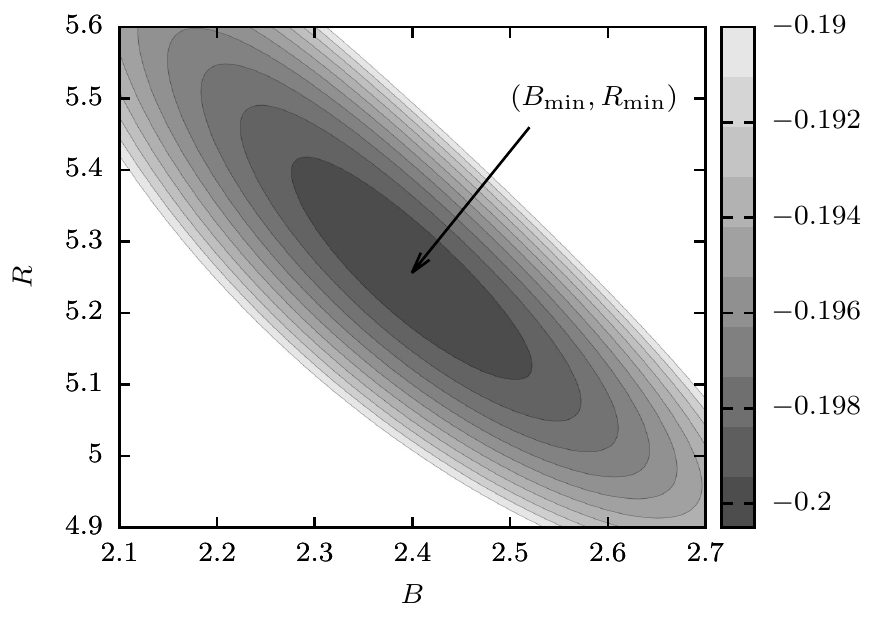}\\
\parbox{0.48\textwidth}{
\caption{Dependence of solutions of Eq.~\eqref{gluon_quasizero_modes_equation} on $BR^2$.\label{lambdas_figure}}
}
\hfill
\parbox{0.48\textwidth}{
\caption{Free energy density $\delta F_\varkappa(B,R)$ given by Eq.~\eqref{effpot_total_kappa} with $\varkappa$ given by Eq.~\eqref{kappa} ($z_0=100,a=10$).\label{effpot_total_figure2}}
}
\end{figure}

As it is shown above, a ``mass''  $\varkappa$ for quasi-zero modes  is required to reproduce correct strong-field asymptotics of the free energy density in the infinite volume limit.  However, in the finite volume $\varkappa$ should depend on the size of the domain via dimensionless parameter $z=BR^2$. In the previous section calculations were performed with constant value $\varkappa$ in agreement with the infinite volume result.  
At small $R$ and $B$, eigenvalues for all gluon and ghost modes (including quasi-zero modes) quickly increase Dependence of gluon eigenvalues on $BR^2$ is illustrated by Fig.~\ref{lambdas_figure}. Given that effective ``mass'' accumulates contributions proportional to the inverse eigenvalues, one may expect  that the effective ``mass'' vanishes at small  $z$.  Therefore, it is plausible that the effective ``mass'' satisfies conditions
\begin{equation*}
\varkappa(z)\xrightarrow{z\to 0}0,
\quad
\varkappa(z)\xrightarrow{z\to\infty}\varkappa\not= 0.
\end{equation*}
Depending on the profile of the function $\varkappa(z)$, there may exist global minimum of the free energy density with respect to both background field strength $B$ and domain size $R$, which would stabilize the mean size of domains in the ensemble of almost everywhere homogeneous Abelian (anti-)self-dual fields. 
Free energy density $\delta F_\varkappa(B,R)$ with a global minimum in $R$ and $B$ calculated for  sample $\varkappa(z)$,
\begin{equation}
\label{kappa}
\varkappa(z)=\frac{2}{\pi}\left[\arctan\exp\left(\frac{z_0-z}{a}\right)+\arctan\exp\left(\frac{z-z_0}{a}\right)-2\arctan\exp\left(-\frac{z_0}{a}\right)\right],
\end{equation}
is shown in Fig.\ref{effpot_total_figure2}.

To summarize, it becomes clear that explicit calculation of the finite-size dependence of effective quasi-zero ``mass''  is an important task to be done. Another interesting and important task relates to incorporation of the quark contribution to the free energy density. 
Quark field in the presence of self-dual gauge field also has  (quasi-)zero  modes,  and their interplay with gluon modes seems to be important for studying the finite size effects in free energy density.  

\section{Acknowledgements}
We acknowledge fruitful discussions with M. Bordag and I.G. Pirozhenko.

\appendix
\section{Zeta function for ghost and gluon fields\label{ghost_gluon_detailed}}

\subsection{Ghosts}
Here we describe calculation of zeta functions in more detail.
We start with the expression for $\zeta^\text{gh}(s)$ given by~\eqref{ghosts_zeta_start}.
Using formula~\eqref{Kummer_asymptotic_expansion} to re-expand $\Psi^\mathrm{gh}$ in powers of $k^{-1}$ and summing over $m$, one arrives at the following expressions for $u_i^\mathrm{gh}$:
\begin{gather*}
u_0^\mathrm{gh}(t,B,R)=B^2R^4\frac{R^4t^4-2R^2t^2-4+4\left(1+R^2t^2\right)^{1/2}}{24R^4 t^4\left(1+R^2t^2\right)^{1/2}},\\
u_1^\mathrm{gh}(t,B,R)=-B^2R^4\frac{1+3R^2t^2}{48\left(1+R^2t^2\right)},\\
\begin{split}
&u_2^\mathrm{gh}(t,B,R)=\frac{B^2R^4}{192\left(1+R^2t^2\right)^{1/2}}\left[\frac{16}{R^2t^2}+\frac{R^2t^2(32+7R^2t^2)}{(1+R^2t^2)^3}+32\frac{1-\left(1+R^2t^2\right)^{1/2}}{R^4t^4}\right]\\
&+\frac{B^4R^8}{1920\left(1+R^2t^2\right)^{1/2}}
\left[\frac{32}{R^4t^4}-\frac{128}{R^6t^6}-\frac{16}{R^2t^2}+\frac{10+13R^2t^2}{(1+R^2t^2)^2}-256\frac{1-\left(1+R^2t^2\right)^{1/2}}{R^8t^8}\right].
\end{split}
\end{gather*}
The sums over $k$ are calculated for $\mathfrak{R}s>1$ and analytically continued in terms of Riemann $\zeta$ function
\begin{equation*}
\sum_{k=1}^\infty \frac{k^{1-2s}}{k^i}=\zeta(i-1+2s)
\end{equation*}
to the strip $0<\mathfrak{R}s<1$ where integrals over $t$ converge at $t\to 0$.
One obtains
\begin{multline*}
\sum_{k=1}^\infty k^{1-2s} \int_0^\infty \frac{dt}{t^{2s}}\frac{d}{dt}\sum_{i=0}^2 \frac{u_i^{\mathrm{gh}}(t,B,R)}{k^i}=
B^2R^{4+2s}\left\{\vphantom{\frac11}\right.\zeta(-1+2s)\frac{-\Gamma(2-s)\Gamma(1/2+s)}{24\sqrt{\pi}(2+s)}+\\
\left. \zeta(2s)\frac{\pi s(1-2s)}{48\sin\pi s}+
\zeta(1+2s)\frac{\left(6-s(2+s)(11+5s)\right)\Gamma(1-s)\Gamma(3/2+s)}{72\sqrt{\pi}(2+s)}
\right\}
-B^4R^{8+2s}\zeta(1+2s)\frac{s\Gamma(2-s)\Gamma(3/2+s)}{480\sqrt{\pi}(4+s)}.
\end{multline*}
The expansion of counterterms in powers of $s$ around $s=0$ is
\begin{multline*}
\frac{\sin\pi s}{\pi}\sum_{k=1}^\infty k^{1-2s} \int_0^\infty \frac{dt}{t^{2s}}\frac{d}{dt}\sum_{i=0}^2 \frac{u_i^{\mathrm{gh}}(t,B,R)}{k^i}=\\
B^2R^4\left[-\frac{1}{96}+\frac{1}{288}(4-6\gamma+\log 64-6\log R)s\right]+
B^4R^8\left[-\frac{1}{7680}s\right]+O(s^2),
\end{multline*}
where $\gamma=0.5772156649\dots$ is Euler's constant, and
\begin{equation*}
\mathrm{Tr}\breve{n}^2=3,\quad \mathrm{Tr}\breve{n}^4=4\frac{9}{16}=\frac{9}{4}.
\end{equation*}
Since the whole spectrum is invariant with respect to $B\to -B$, trace over color leads to factor 4. Evaluating the derivative of $\zeta^\mathrm{gh}(s)$ with respect to $s$ at $s=0$, one arrives at the expression~\eqref{ghosts_effpot}.

\subsection{Gluons}

The same procedure is applied to zeta function of gluons.
With Dirichlet boundary condition for color-charged modes, the whole set of eigenvalues is determined by the equations~\cite{Kalloniatis:2001dw}
\begin{gather*}
M\left(\frac{k}{2}+1-m\pm 1-\frac{\lambda^2}{2n_a B},k+2,\frac{n_a BR^2}{2}\right)=0,\quad
k=0,1,\dots\quad m=-\frac{k}{2},\dots,\frac{k}{2},
\end{gather*}
and every solution of these equations with given $k,m$ and $a$ is $2(k+1)$-degenerate. Repeating the procedure carried out for ghosts, one finds
\begin{multline}
\label{zeta_gl}
\zeta^\text{gl}(s)=\mathrm{Tr}_\mathrm{c}\frac{\sin \pi s}{\pi}\sum_{k=0}^\infty \sum_{m=-\frac{k}{2}}^\frac{k}{2}2(k+1) \int_0^\infty \frac{dt}{t^{2s}}\frac{d}{dt}\Psi^\mathrm{gl}(k+1,m,t,\breve{B},R)=\\
2\mathrm{Tr}_\mathrm{c}\left\{\frac{\sin\pi s}{\pi}\sum_{k=1}^\infty k^{1-2s} \int_0^\infty \frac{dt}{t^{2s}}\frac{d}{dt}\left[\sum_{m=-\frac{k-1}{2}}^\frac{k-1}{2}\Psi^\mathrm{gl}(k,m,kt ,\breve{B},R)-\sum_{i=0}^2  \frac{u_i^{\mathrm{gl}}(t,\breve{B},R)}{k^i}\right]\right.\\
\left.
+\frac{\sin\pi s}{\pi}\sum_{k=1}^\infty k^{1-2s} \int_0^\infty \frac{dt}{t^{2s}}\frac{d}{dt}\sum_{i=0}^2 \frac{u_i^{\mathrm{gl}}(t,\breve{B},R)}{k^i},
\right\}
\end{multline}
where
\begin{equation*}
\begin{split}
\Psi^\mathrm{gl}(k,m,t,B,R)=
\log\frac{\exp\left(-\frac{BR^2}{2}\right)M\left(\frac{k+1}{2}-m+1+\frac{t^2}{2B},k+1,\frac{BR^2}{2}\right)M\left(\frac{k+1}{2}-m-1+\frac{t^2}{2B},k+1,\frac{BR^2}{2}\right)}{\left(k!\left(\frac{t R}{2}\right)^{-k}I_{k}(t R)\right)^2}.
\end{split}
\end{equation*}
Coefficients of asymptotic expansion $u_i^\mathrm{gl}$ are given by
\begin{gather*}
u_0^\mathrm{gl}(t,B,R)=B^2R^4\frac{R^4t^4-2R^2t^2-4+4\left(1+R^2t^2\right)^{1/2}}{12R^4 t^4\left(1+R^2t^2\right)^{1/2}},\\
u_1^\mathrm{gl}(t,B,R)=-B^2R^4\frac{1+3R^2t^2}{24\left(1+R^2t^2\right)},\\
\begin{split}
&u_2^\mathrm{gl}(t,B,R)=\frac{B^2R^4}{96\left(1+R^2t^2\right)^{1/2}}\left[-\frac{80}{R^2t^2}+\frac{R^2t^2(32+7R^2t^2)}{(1+R^2t^2)^3}-160\frac{1-\left(1+R^2t^2\right)^{1/2}}{R^4t^4}\right]\\
&+\frac{B^4R^8}{960\left(1+R^2t^2\right)^{1/2}}
\left[-\frac{128}{R^6t^6}+\frac{32}{R^4t^4}-\frac{16}{R^2t^2}+\frac{10+13R^2t^2}{(1+R^2t^2)^2}-256\frac{1-\left(1+R^2t^2\right)^{1/2}}{R^8t^8}\right].
\end{split}
\end{gather*}
And one obtains the following analytical continuation of the counterterms
\begin{multline*}
\sum_{k=1}^\infty k^{1-2s} \int_0^\infty \frac{dt}{t^{2s}}\frac{d}{dt}\sum_{i=0}^2 \frac{u_i^{\mathrm{gl}}(t,B,R)}{k^i}=
B^2R^{4+2s}\left\{\vphantom{\frac11}\right.\zeta(-1+2s)\frac{-\Gamma(2-s)\Gamma(1/2+s)}{12\sqrt{\pi}(2+s)}+
\zeta(2s)\frac{\pi s(1-2s)}{24\sin\pi s}+
\\\left.
\zeta(1+2s)\frac{\left(30+s(2+s)(11+5s)\right)\Gamma(1-s)\Gamma(3/2+s)}{36\sqrt{\pi}(2+s)}
\right\}
-B^4R^{8+2s}\zeta(1+2s)\frac{s\Gamma(2-s)\Gamma(3/2+s)}{240\sqrt{\pi}(4+s)}.
\end{multline*}
Now, counterterms can be expanded in powers of $s$:
\begin{multline*}
\frac{\sin\pi s}{\pi}\sum_{k=1}^\infty k^{1-2s} \int_0^\infty \frac{dt}{t^{2s}}\frac{d}{dt}\sum_{i=0}^2 \frac{u_i^{\mathrm{gl}}(t,B,R)}{k^i}=\\
B^2R^4\left[\frac{5}{48}+\frac{1}{144}(31+30\gamma-30\log 2+30\log R)s\right]+
B^4R^8\left[-\frac{1}{3840}s\right]+O(s^2).
\end{multline*}
Finally,
\begin{multline*}
\delta U^\mathrm{gl}(B,R)=-\frac{1}{2}\left.\frac{d}{ds}\zeta^\text{gl}(s)\right|_{s=0}=
4\sum_{k=1}^\infty k \left[\sum_{m=-\frac{k-1}{2}}^\frac{k-1}{2}\Psi^\text{gl}(k,m,0,\frac{\sqrt{3}B}{2},R)-\frac{3}{4}B^2R^4\frac{1}{24}\left(1-\frac{1}{k}-\frac{5}{k^2}\right)\right]\\
-\frac{B^2R^4}{48}(31+30\gamma-30\log 2+30\log R)+\frac{3B^4R^8}{5120}.
\end{multline*}

\subsection{Contribution of quasi-zero modes}
As it is explained in Section~\ref{section_ghosts_and_gluons}, simple one-loop expression for $\delta U^\mathrm{gl}(B,R)$ has incorrect asymptotic behavior versus $B$ and $R$ due to quasi-zero modes of gluons.
Now, let us calculate the contribution of quasi-zero modes $\lambda_{k\frac{k}{2}0}$ to the effective potential. These modes are given by lowest solutions of the equations
\begin{equation}
\label{gluon_quasizero_modes_equation}
M\left(-\frac{\lambda^2}{2|\breve{B}|},k+2,\frac{|\breve{B}|R^2}{2}\right)=0,\quad
k=0,1,2,\dots
\end{equation}
that reduce to
\begin{equation*}
\left(\frac{\lambda R}{2}\right)^{-k-1}J_{k+1}(\lambda R)=0,\quad k=0,1,\dots
\end{equation*}
at $B\to 0$.
The contribution of quasi-zero modes to the effective potential can be expressed as
\begin{gather}
\nonumber
\delta U^{\textrm{gl}(0)}(B,R)=\frac{1}{2}\textrm{Tr} \log\frac{\lambda_{k\frac{k}{2}0}^2(B,R)}{\lambda_{k0}^2(0,R)}=\left.-\frac{1}{2}\frac{d}{ds}\zeta^{\textrm{gl}(0)}(s)\right|_{s=0},\\
\label{zeta_quasi-zero_gl}
\zeta^{\textrm{gl}(0)}(s)=2\mathrm{Tr}_\mathrm{c}\sum_{k=0}^\infty (k+1) \left( \lambda_{k\frac{k}{2}0}^{-2s}(B,R)-\lambda_{k0}^{-2s}(0,R) \right),
\end{gather}
where factor $2$ in the definition of $\zeta^{\textrm{gl}(0)}(s)$ originates from two polarizations of quasi-zero gluon modes. For the sake of brevity we omit color eigenvalue and restore it in the final answer ($B\to \frac{\sqrt{3}}{2}B$ for adjoint representation of $su(3)$).

To continue $\zeta(s)$ to $s\to 0$, we add and subtract several terms of asymptotic expansion in $k\gg 1$ of $d\zeta/dB$ and  $d^2\zeta/dB^2$ at $B=0$. $\lambda(B)$ is given implicitly via Eq.~\eqref{gluon_quasizero_modes_equation}, and we find derivatives
\begin{gather*}
\left.\frac{d\lambda_{k\frac{k}{2}0}(B,R)}{dB}\right|_{B=0}=-\frac{k+2}{2\lambda_{k0}(0,R)},\ 
\left.\frac{d^2\lambda_{k\frac{k}{2}0}(B,R)}{dB^2}\right|_{B=0}= -\frac{k^2+8k+12- \lambda_{k0}^2(0,R)R^2}{12 \lambda_{k0}^3(0,R)},
\end{gather*}
where we used identities (\cite{Abramowitz:1972}, 9.1.27)
\begin{gather*}
J_{\nu-1}(z)+J_{\nu+1}=\frac{2\nu}{z}J_\nu(z),\quad
J'_\nu(z)=-J_{\nu+1}(z)+\frac{\nu}{z}J_\nu(z)
\end{gather*}
and the fact that
\begin{equation*}
J_{k+1}(\lambda_{k0}(0,R)R)=0.
\end{equation*}
We obtain
\begin{gather}
\label{gl_first_derivative}
\begin{split}
\left.\frac{d\zeta^{\textrm{gl}(0)}}{dB}\right|_{B=0}&=8\sum_{k=0}^\infty (k+1)(-2s)\lambda_{k\frac{k}{2}0}^{-2s-1}(B,R)\left.\frac{d\lambda_{k\frac{k}{2}0}(B,R)}{dB}\right|_{B=0}\\
&=8\sum_{k=0}^\infty (k+1)(-2s)\lambda_{k\frac{k}{2}0}^{-2s-1}(0,R)\left(-\frac{k+2}{2\lambda_{k0}(0,R)}\right)=8\sum_{k=1}^\infty \zeta_k^{(1)},
\end{split}\\
\label{gl_second_derivative}
\begin{split}
\left.\frac{d^2\zeta^{\textrm{gl}(0)}}{dB^2}\right|_{B=0}=&8\sum_{k=0}^\infty (k+1)\left[(-2s)(-2s-1)\lambda_{k\frac{k}{2}0}^{-2s-2}(0,R)\left(-\frac{k+2}{2\lambda_{k0}(0,R)} \right)^2\right.\\
&\left.-2s\lambda_{k\frac{k}{2}0}^{-2s-1}(0,R)\left(-\frac{k^2+8k+12-\lambda_{k0}(0,R)R^2}{12 \lambda_{k0}^3(0,R)}\right)\right]=8\sum_{k=1}^\infty \zeta_k^{(2)},
\end{split}
\end{gather}
where additional factor $4$ is due to color trace. Note that summation index is shifted $k\to k-1$ in the definition of $\zeta^{(i)}_k$. Next, we use uniform asymptotic expansion of zeros of Bessel functions (see \cite{DLMF}, Eq.~10.21.vii)
\begin{equation*}
\rho_\nu(t)=\nu \sum_{k=0}^\infty\frac{\alpha_k}{\nu^{2k/3}},\quad \theta\left(-2^\frac{1}{3}\alpha\right)=\pi t,\quad
\alpha_0=1,\quad \alpha_1=\alpha,\dots,
\end{equation*}
where $\theta(x)$ is the phase of Airy functions
\begin{equation*}
\theta(x)=\arctan\frac{\mathrm{Ai}(x)}{\mathrm{Bi}(x)}.
\end{equation*}
With $t=1$ this formula gives the desired expansion for the first zero $z\neq 0$ of $J_\nu(z)$, where $\alpha=1.855757\dots$ Substituting this expansion into Eqs.~\eqref{gl_first_derivative} and \eqref{gl_second_derivative} and re-expanding in powers of $k^{-1}$, we find
\begin{gather*}
\zeta_k^{(1)}=R^{2+2s}s\left[k^{-2s}-2\alpha (1+s)k^{-2/3-2s}+k^{-1-2s}\right]+O\left(k^{-4/3-2s}\right)=
Z_k^{(1)}+O\left(k^{-4/3-2s}\right),
\\
\zeta_k^{(2)}=R^{4+2s}k^{-1-2s}\frac{s}{2}(1+2s)+O\left(k^{-5/3-2s}\right)=Z_k^{(2)}+O\left(k^{-5/3-2s}\right).
\end{gather*}
Next, we split $\zeta^{\textrm{gl}(0)}$ into two parts
\begin{equation*}
\zeta^{\textrm{gl}(0)}(s)=8\left\{\sum_{k=0}^\infty \left[(k+1) \left( \lambda_{k\frac{k}{2}0}^{-2s}(B,R)-\lambda_{k0}^{-2s}(0,R) \right)-BZ_{k+1}^{(1)}-\frac{B^2}{2}Z_{k+1}^{(2)}\right]
+B\sum_{k=0}^\infty Z_{k+1}^{(1)}+\frac{B^2}{2} \sum_{k=0}^\infty Z_{k+1}^{(2)}\right\}.
\end{equation*}
The first sum is an analytic function for $\mathfrak{R} s>0$. The remaining sums are evaluated for $\mathfrak{R}s>1/2$ and analytically continued to $s\to 0$: 
\begin{gather*}
Z^{(1)}=\sum_{k=1}^\infty Z_k^{(1)}=R^{2+2s}s\left[\zeta(2s)-2\alpha (1+s)\zeta\left(\frac{2}{3}+2s\right)+\zeta(1+2s)\right],\\
Z^{(2)}=\sum_{k=1}^\infty Z_k^{(2)}=R^{4+2s}\frac{s}{2}(1+2s)\zeta(1+2s).
\end{gather*}
Finally, we obtain
\begin{multline*}
\delta U^{\textrm{gl}(0)}(B,R)=-\frac{1}{2}\left.\frac{d}{ds}\zeta^{\textrm{gl}(0)}(s)\right|_{s=0}=\\
-4\sum_{k=0}^\infty \left[-(k+1)\log\frac{\lambda_{k\frac{k}{2}0}^2(B,R)}{\lambda_{k0}^2(0,R)}-BR^2\left(1-2\alpha (k+1)^{-2/3}+(k+1)^{-1}\right)-\frac{B^2 R^4}{4(k+1)}\right]\\
-2BR^2\left[-1+2\gamma+2\log R-4\alpha\zeta\left(\frac23\right)\right]-B^2R^4(1+\gamma+\log R).
\end{multline*}

\subsection{Contribution of quasi-zero modes with effective ``mass''}

If one includes ``mass term'' $\varkappa$ for quasi-zero modes, the formulas for the effective potential become
\begin{gather}
\nonumber
\delta U_\varkappa^{\textrm{gl}(0)}(B,R)=\frac{1}{2}\textrm{Tr} \log\frac{\lambda_{k\frac{k}{2}0}^2(B,R)+\varkappa B}{\lambda_{k0}^2(0,R)}=\left.-\frac{1}{2}\frac{d}{ds}\zeta^{\textrm{gl}(0)}(s)\right|_{s=0},\\
\label{zeta_quasi-zero_gl_with_kappa}
\zeta_\varkappa^{\textrm{gl}(0)}(s)=2\mathrm{Tr}_\mathrm{c}\sum_{k=0}^\infty (k+1) \left( \left(\lambda_{k\frac{k}{2}0}^2(B,R)+\varkappa B\right)^{-s}-\lambda_{k0}^{-2s}(0,R) \right).
\end{gather}
In complete analogy to contribution without $\varkappa$,
\begin{equation*}
\left.\frac{d\zeta_\varkappa^{\textrm{gl}(0)}}{dB}\right|_{B=0}=8\sum_{k=1}^\infty \zeta_{\varkappa k}^{(1)},
\quad
\left.\frac{d^2\zeta_\varkappa^{\textrm{gl}(0)}}{dB^2}\right|_{B=0}=8\sum_{k=1}^\infty \zeta_{\varkappa k}^{(2)}.
\end{equation*}
The corresponding asymptotic expansions in powers of $k$ are
\begin{gather*}
\zeta_{\varkappa k}^{(1)}=R^{2+2s}s\left[k^{-2s}-2\alpha (1+s)k^{-2/3-2s}+\left(1-\varkappa \right)k^{-1-2s}\right]+O\left(k^{-4/3-2s}\right)=
Z_{\varkappa k}^{(1)}+O\left(k^{-4/3-2s}\right),
\\
\zeta_{\varkappa k}^{(2)}=R^{4+2s}k^{-1-2s}\frac{s}{2}\left(1+2s\right)+O\left(k^{-5/3-2s}\right)=Z_{\varkappa k}^{(2)}+O\left(k^{-5/3-2s}\right).
\end{gather*}
Counterterms are summed for $\mathfrak{R} s>1/2$
\begin{gather*}
Z_\varkappa^{(1)}=\sum_{k=1}^\infty Z_{\varkappa k}^{(1)}=R^{2+2s}s\left[\zeta(2s)-2\alpha (1+s)\zeta\left(\frac{2}{3}+2s\right)+\left(1-\varkappa\right)\zeta(1+2s)\right],\\
Z_\varkappa^{(2)}=\sum_{k=1}^\infty Z_{\varkappa k}^{(2)}=R^{4+2s}\frac{s}{2}\left(1+2s\right)\zeta(1+2s).
\end{gather*}
Finally,
\begin{multline*}
\delta U_\varkappa^{\textrm{gl}(0)}(B,R)=\\
-4\sum_{k=0}^\infty \left[-(k+1)\log\frac{\lambda_{k\frac{k}{2}0}^2(B,R)+\varkappa B}{\lambda_{k0}^2(0,R)}-BR^2\left(1-2\alpha (k+1)^{-2/3}+(1-\varkappa)(k+1)^{-1}\right)-\frac{B^2 R^4}{4(k+1)}\right]\\
-2BR^2\left[-1+(2\gamma+2\log R)(1-\varkappa)-4\alpha\zeta\left(\frac23\right)\right]-B^2R^4\left[1+(\gamma+\log R)\right]
\end{multline*}

\subsection{Contribution of all eigenmodes with effective ``mass'' for quasi-zero modes}

Now we are ready to incorporate $\varkappa$ for quasi-zero modes into one-loop effective potential.
The desired zeta function is written as
\begin{equation*}
\zeta_\varkappa^\mathrm{gl}(s)=\zeta^\mathrm{gl}(s)-\zeta^{\textrm{gl}(0)}(s)+\zeta_\varkappa^{\textrm{gl}(0)}(s),
\end{equation*}
where zeta functions in right-hand side are given by Eqs.~\eqref{zeta_gl},\eqref{zeta_quasi-zero_gl} and~\eqref{zeta_quasi-zero_gl_with_kappa}. The corresponding effective potential is given by
\begin{multline*}
\delta U_\varkappa^\mathrm{gl}(B,R)=
4\sum_{k=1}^\infty k \left[\sum_{m=-\frac{k-1}{2}}^\frac{k-1}{2}\Psi^\text{gl}(k,m,0,\frac{\sqrt{3}B}{2},R)-\frac{3}{4}B^2R^4\frac{1}{24}\left(1-\frac{1}{k}-\frac{5}{k^2}\right)\right]\\
-\frac{B^2R^4}{48}(31+30\gamma-30\log 2+30\log R)+\frac{3B^4R^8}{5120}\\
+4\sum_{k=0}^\infty \left[(k+1)\log\frac{\lambda_{k\frac{k}{2}0}^2(B,R)+\varkappa B}{\lambda_{k\frac{k}{2}0}^2(B,R)}-\varkappa BR^2(k+1)^{-1}\right]
+4\varkappa BR^2(\gamma+\log R).
\end{multline*}
Thus obtained free energy is not an even function of $B$. To restore this property, one adds term $\varkappa$ to contribution of modes $\lambda_{k,-\frac{k}{2},0}^2(B,R)$, and beta function of $SU(3)$ gluodynamics emerges.
After these steps one obtains formula~\eqref{gluon_effpot_mass} for free energy.

\section{Connection between Kummer and Bessel functions\label{Kummer_functions}}
We find the desired limit from definition of Kummer function via series (\cite{Abramowitz:1972}, 13.1.2)
\begin{equation*}
\lim_{z\to 0} M\left(a+\frac{b}{z},c,dz\right)=\lim_{z\to 0} \sum_{n=0}^\infty \frac{\left(a+\frac{b}{z}\right)_n(dz)^n}{(c)_n n!}=\sum_{n=0}^\infty \frac{(bd)^n}{(c)_n n!}={_0F_1}(c,bd)
\end{equation*}
$(c)_n$ is Pochhammer symbol:
\begin{equation*}
(c)_0=1,\quad (c)_n=c(c+1)\cdots(c+n-1).
\end{equation*}
Comparing series expansions of Bessel functions (\cite{Abramowitz:1972}, 9.1.10, 9.6.10) and hypergeomteric function ${}_0F_1$, we find
\begin{gather*}
{_0F_1}(a,z)=\sum_{k=0}^\infty \frac{z^k \Gamma(a)}{k! \Gamma(a+k)}=
(a-1)!\left(\sqrt{z}\right)^{1-a} \left(\sqrt{z}\right)^{a-1} \sum_{k=0}^\infty \frac{(\sqrt{z})^{2k} }{k! \Gamma(a-1+k+1)}=(a-1)! \left(\sqrt{z}\right)^{1-a} I_{a-1}(2\sqrt{z}),\\
{_0F_1}(a,-z)=\sum_{k=0}^\infty \frac{(-z)^k \Gamma(a)}{k! \Gamma(a+k)}=
(a-1)!\left(\sqrt{z}\right)^{1-a} \left(\sqrt{z}\right)^{a-1} \sum_{k=0}^\infty \frac{(-1)^k(\sqrt{z})^{2k} }{k! \Gamma(a-1+k+1)}=(a-1)! \left(\sqrt{z}\right)^{1-a} J_{a-1}(2\sqrt{z})
\end{gather*}
for $z\geqslant 0$.

\section{Asymptotic expansion of $M(\frac{m+1}{2}+\mu-\kappa+n+\mu^2t^2,1+2\mu+m,z)$ as $\mu\to\infty$, $\kappa/\mu$ fixed\label{asymptotic_expansion_appendix}}

We use the method described in~\cite{Olver:1997}. The Whittaker function
\begin{equation*}
M_{\kappa-n-\mu^2t^2,\mu+\frac{m}{2}}(z)=
\exp\left(-\frac{z}{2}\right)z^{\mu+\frac{m}{2}+\frac{1}{2}}M\left(\frac{1}{2}+\mu+\frac{m}{2}-\kappa+n+\mu^2t^2,1+2\mu+m,z\right)
\end{equation*}
satisfies equation
\begin{gather*}
\frac{d^2M_{\kappa-n-\mu^2t^2,\mu+\frac{m}{2}}}{dz^2}=
\left(\frac{1}{4}+\frac{n}{z}+\frac{m^2-1}{4z^2}+2\mu\left(-\frac{\kappa}{2\mu}+\frac{m}{2z^2}\right)+4\mu^2\left(\frac{t^2}{4z}+\frac{1}{4z^2}\right)\right)M_{\kappa-n-\mu^2t^2,\mu+\frac{m}{2}},
\\
f_0(z)=\frac{t^2}{4z}+\frac{1}{4z^2},\quad f_1(z)=-\frac{\kappa}{2\mu}+\frac{m}{2z^2},\quad f_2(z)=\frac{1}{4}+\frac{n}{z}+\frac{m^2-1}{4z^2}.
\end{gather*}
Formal solution to the above equation is sought in the form
\begin{gather*}
M_{\kappa-n-\mu^2t^2,\mu+\frac{m}{2}}=f_0^{-\frac{1}{4}}(z)\exp\left(2\mu\xi(z)\right)X(z)\sum_{s=0}^\infty \frac{A_s(\xi(z))}{(2\mu)^s},
\end{gather*}
where
\begin{gather*}
A_0=1,\quad X(\xi)=\exp\left(\frac{1}{2}\int d\xi \phi\right),\quad \psi(z)=\frac{f_2(z)}{f_0(z)}-f_0^{-\frac{3}{4}}(z)\frac{d^2}{dz^2}f_0^{-\frac{1}{4}}(z).
\\
\xi(z)=\int dz\sqrt{f_0(z)}=\sqrt{1+t^2z}+\log\frac{t\sqrt{z}}{\sqrt{1+t^2z}+1},\quad \phi(z)=\frac{f_1(z)}{f_0(z)},\\
\int d\xi \phi=\int dz \frac{d\xi}{dz}\phi=\int dz \sqrt{f_0(z)}\frac{f_1(z)}{f_0(z)}=\int \frac{dz}{\sqrt{1+t^2z}}\left(-\frac{\kappa}{\mu}+\frac{m}{z}\right)=
-\frac{2\kappa}{\mu t^2}\sqrt{1+t^2z}+2m\log\frac{t\sqrt{z}}{\sqrt{1+t^2z}+1}.
\end{gather*}
The leading asymptotics is
\begin{multline*}
M\left(\frac{m+1}{2}+\mu-\kappa+n+\mu^2t^2,1+2\mu+m,z\right)=\\
\exp\left(\frac{z}{2}\right)z^{-\mu-\frac{m}{2}-\frac{1}{2}}M_{\kappa-n-\mu^2t^2,\mu+\frac{m}{2}}(z)\sim
C\exp\left(\frac{z}{2}\right)z^{-\mu-\frac{m}{2}-\frac{1}{2}}\frac{\sqrt{2z}}{(1+t^2z)^{\frac{1}{4}}}\\
\times \exp\left(2\mu\sqrt{1+t^2z}+2\mu\log\frac{t\sqrt{z}}{\sqrt{1+t^2z}+1}-\frac{\kappa}{\mu t^2}\sqrt{1+t^2z}+m\log\frac{t\sqrt{z}}{\sqrt{1+t^2z}+1}\right).
\end{multline*}
The multiplicative constant $C$ is fixed by the condition
$
M\left(a,b,0\right)=1
$:
\begin{multline}\label{Kummer_asymptotic_expansion}
M\left(\frac{m+1}{2}+\mu-\kappa+n+\mu^2t^2,1+2\mu+m,z\right)\sim
\\
\exp\left(\frac{z}{2}-2\mu\right)\frac{2^{2\mu+m}}{(1+t^2z)^{\frac{1}{4}}}\exp\left(2\mu\sqrt{1+t^2z}-(2\mu+m) \log\left[\sqrt{1+t^2z}+1\right]-\frac{\kappa}{\mu}\frac{z}{\sqrt{1+t^2z}+1}\right)
\sum_{s=0}^\infty \frac{A_i(z)}{(2\mu)^s}.
\end{multline}
The coefficients $A_i(z)$ are found with the help of recursion relation ($s\geqslant 0, A_0=1$)
\begin{gather*}
A_{s+1}(\xi)=-\frac{1}{2}\phi-\frac{1}{2}A_s'+\int d\xi \left(\psi +\frac{1}{2}\phi'-\frac{1}{4}\phi^2\right)A_s,\\
A_{s+1}(\xi)=-\frac{1}{2}\phi-\frac{1}{2}\frac{dA_s}{dz}\frac{dz}{d\xi}+\int dz \frac{d\xi}{dz} \left(\psi +\frac{1}{2}\frac{d\phi}{dz}\frac{dz}{d\xi}-\frac{1}{4}\phi^2\right)A_s,\quad
\frac{d\xi}{dz}=\sqrt{f_0(z)}.
\end{gather*}
The constants of integration are fixed by the requirement
\begin{equation*}
A_{s}(0)=0,\quad s\geqslant 1.
\end{equation*}
Expansion~\eqref{Kummer_asymptotic_expansion} is valid for bounded $z\geqslant 0$.

\bibliography{references}

\end{document}